\documentclass{Interspeech}
\usepackage{float}

\interspeechcameraready

\title{Enhancing Speech Emotion Recognition with Graph-Based Multimodal Fusion and Prosodic Features for the Speech Emotion Recognition in Naturalistic Conditions Challenge at Interspeech 2025}

\author[affiliation={1}]{Alef Iury}{Ferreira}
\author[affiliation={1}]{Lucas Rafael}{Gris}
\author[affiliation={1}]{Alexandre Ferro}{Filho}
\author[affiliation={1}]{Lucas}{Ólives}
\author[affiliation={1}]{Daniel}{Ribeiro}
\author[affiliation={1}]{Luiz}{Fernando}
\author[affiliation={2}]{Fernanda}{Lustosa}
\author[affiliation={3}]{Rodrigo}{Tanaka}
\author[affiliation={4}]{Frederico}{Oliveira}
\author[affiliation={1}]{Arlindo Galvão}{Filho}

\affiliation{}{Federal University of Goiás}{Brazil}
\affiliation{}{Federal University of Rio Grande do Norte}{Brazil}
\affiliation{}{Aeronautics Institute of Technology}{Brazil}
\affiliation{}{Federal University of Mato Grosso}{Brazil}
\email{alef\_iury\_c.c@discente.ufg.br, lucas.gris@discente.ufg.br, fred.santos.oliveira@gmail.com}

\usepackage{comment}
\usepackage{todonotes}
\begin{document}
\maketitle

\vspace{1\baselineskip}
\begin{abstract}
Training SER models in natural, spontaneous speech is especially challenging due to the subtle expression of emotions and the unpredictable nature of real-world audio. In this paper, we present a robust system for the INTERSPEECH 2025 Speech Emotion Recognition in Naturalistic Conditions Challenge, focusing on categorical emotion recognition. Our method combines state-of-the-art audio models with text features enriched by prosodic and spectral cues. In particular, we investigate the effectiveness of Fundamental Frequency (F0) quantization and the use of a pretrained audio tagging model. We also employ an ensemble model to improve robustness. On the official test set, our system achieved a Macro F1-score of 39.79\% (42.20\% on validation). Our results underscore the potential of these methods, and analysis of fusion techniques confirmed the effectiveness of Graph Attention Networks. Our source code is publicly available.\footnote{\url{https://github.com/alefiury/InterSpeech-SER-2025}}.
\end{abstract}

\section{Introduction}

Speech emotion recognition (SER) is a critical component of affective computing and human–computer interaction~\cite{picard1997affective}. Early SER relied on hand-crafted features but struggled with real-world generalization~\cite{busso2008iemocap}. Several factors contribute to these limitations, including individual variability in emotional expression and the subjective nature of emotion labeling~\cite{8682170}. To address these challenges, recent advances in deep learning and self-supervised learning (SSL) have transformed the field by leveraging large-scale unlabeled speech data to learn robust and transferable representations, significantly enhancing SER performance~\cite{morais2022speech}. However, challenges remain when processing natural speech because emotions are expressed subtly and in a spontaneous way~\cite{chakraborty2017analyzing}. 

Recent work has shown that multi-task learning can improve SER performance by jointly addressing related tasks such as speech-to-text recognition and emotion classification~\cite{cai21b_interspeech}. Other studies have introduced domain-specific pretext tasks that use audiovisual cues to learn better emotional representations~\cite{goncalves22_interspeech}. 
In the Odyssey 2024 Speech Emotion Recognition Challenge~\cite{goncalves2024odyssey}, several studies adopted a multimodal approach by combining audio and text inputs, achieving significantly better performance compared to audio-only methods. In this context, the integration of SSL models in a multimodal context has emerged as a promising approach in the field.  

In this paper, we present a system for the INTERSPEECH 2025 Speech Emotion Recognition in the Naturalistic Conditions Challenge. The challenge consists of two tracks: categorical emotion recognition and emotional attribute prediction. This study focuses on categorical emotion recognition. Our approach builds on recent advances by combining state-of-the-art audio models, including Wav2Vec2~\cite{baevski2020wav2vec}, HuBERT~\cite{hsu2021hubert}, WavLM~\cite{chen2022wavlm}, Whisper~\cite{radford2023robust}, and XEUS~\cite{chen2024towards}, alongside a text encoder based on RoBERTa~\cite{zhuang-etal-2021-robustly}. The feature set is enriched by adding prosodic features like Fundamental Frequency (F0) and spectral features such as Mel-Spectrograms. We also use Sequential Feature Resampling (SeqAug)~\cite{georgiou2023seqaug} for data augmentation and a majority voting ensemble to improve overall performance. We tested several configurations of the system and combined various approaches in a final ensemble model.

Extensive experiments and ablation studies confirm that our bimodal and multimodal fusion techniques significantly improve performance. Our results align with recent research
~\cite{chen24odyssey, goncalves2024odyssey, zaidi2024enhancing}, 
offering valuable insights for the advancement of multimodal emotion recognition. 
Additionally, we also investigate the F0 quantization for prosodic modeling and the use of Consistent Ensemble Distillation (CED)~\cite{dinkel2024ced} for Mel-Spectrogram processing.
The remainder of this paper is structured as follows: Section \ref{sec:related} reviews related work and Section \ref{sec:proposed} outlines the proposed system. Section \ref{sec:exp} details the experimental setup, while Section \ref{sec:res} presents and discusses the results. Finally, Section \ref{sec:conc} concludes the paper.

\begin{figure*}[ht!]
  \centering
  \includegraphics[width=1\textwidth]{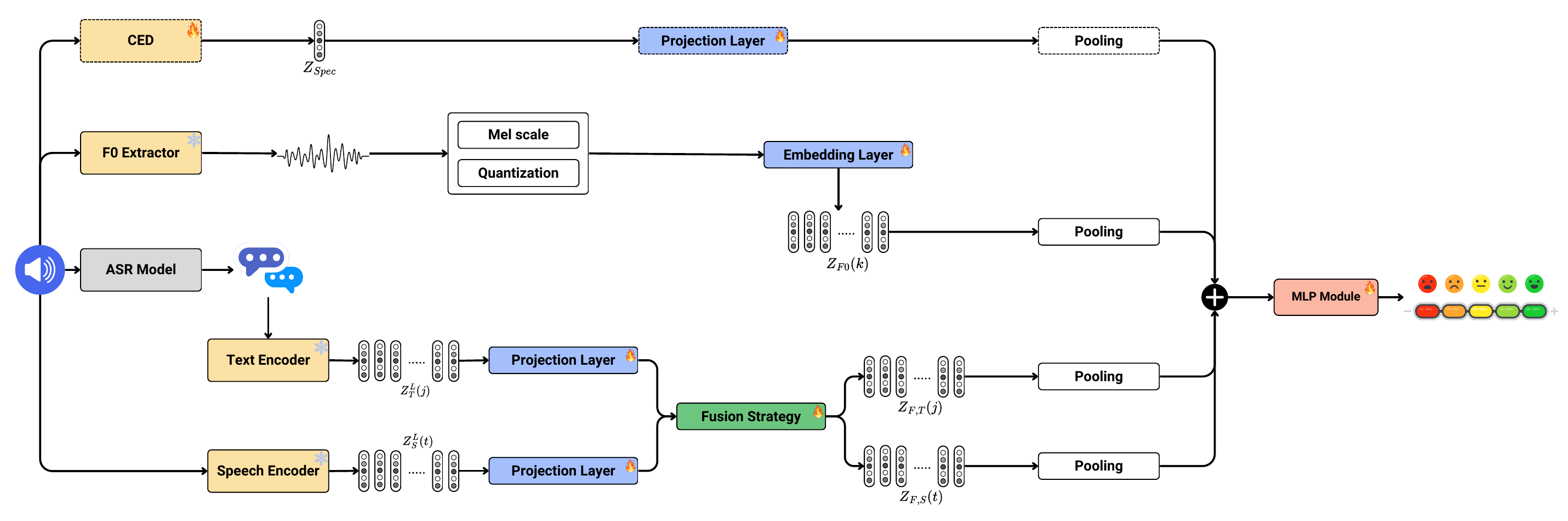}
  \caption{Overview of the proposed system architecture for multimodal emotion recognition. }
\label{fig:proposed-system}
\end{figure*}

\section{Related Work} \label{sec:related}

Recent research has improved SER performance through multimodal fusion, self-supervised learning (SSL), and deep learning architectures. The Odyssey 2024 Challenge demonstrated the advantages of combining speech and text, with top systems leveraging ensemble SSL models, attention-based fusion, and class imbalance mitigation~\cite{goncalves2024odyssey, chen24odyssey}.

Other architectures based on hierarchical cross-attention models (HCAM)~\cite{dutta2023hcam} and Multimodal Dual Attention Transformers (MDAT)~\cite{zaidi2024enhancing} improve feature interaction. HCAM applies bidirectional gated recurrent units (GRUs) with self-attention and cross-modal attention to refine multimodal embeddings, whereas MDAT employs a transformer-based dual-attention mechanism, incorporating both graph attention and co-attention layers to enhance cross-modal dependencies. 

Originally introduced for graph-structured data~\cite{velivckovic2018graph}, Graph Attention Networks (GATs) dynamically assign attention weights to different nodes, allowing for adaptive feature importance estimation. In~\cite{zaidi2024enhancing}, the MDAT integrates graph attention and co-attention mechanisms to enhance multimodal fusion in cross-language SER. Their model applies graph attention layers to dynamically learn dependencies between speech and text embeddings, refining modality-specific features while capturing cross-modal interactions. Additionally, they introduced a transformer encoder layer for high-level feature representation, further improving emotion classification accuracy across multiple languages. Their findings highlight the robustness of graph-based fusion, demonstrating superior generalization compared to conventional multimodal approaches. Our work extends this investigation by evaluating the effectiveness of GATs for multimodal SER in spontaneous speech settings.



\section{Proposed System} \label{sec:proposed}


Figure~\ref{fig:proposed-system} illustrates our system. The hidden states of the last layer \( L \) of the text encoder are denoted by \( Z_T^L(j) \) for positions \( j = 1, \dots, N \) and from the speech encoder by \( Z_S^L(t) \) for frames \( t = 1, \dots, T \). After fusion, modality-specific representations are maintained as \( Z_{F,T}(j) \) (text) and \( Z_{F,S}(t) \) (speech). We further integrate discretized F0 embeddings \( Z_{F0}(k) \) (with \( k \) indexing the F0 sequence) and spectral features \( Z_{Spec} \) from the CED model. Our approach was built based on four main stages:
\begin{enumerate} 
    \item \textbf{Unimodal Speech Model Selection}: We compare several SSL-based audio encoders to identify the most robust unimodal backbone. 
    \item \textbf{Bimodal Fusion with Text}: We incorporate text representations from a pretrained text encoder to leverage content-based features. We also evaluate various fusion strategies.
    \item \textbf{Prosodic and Spectral Feature Integration}: We augment the fused embeddings with discretized F0 embeddings and spectral information through the CED).
    \item \textbf{Additional Strategies}: We investigate some additional strategies in order to enhance the model robustness such as data augmentation and the use of SwiGLU~\cite{shazeer2020glu} activation in the MLP module. 
\end{enumerate} 





\subsection{Unimodal Speech Model Selection} \label{sec:audio_encoders}

We began by comparing five popular open-source SSL speech models: Wav2vec2 Large\footnote{\url{https://huggingface.co/facebook/wav2vec2-large}}, Whisper Large V3\footnote{\url{https://huggingface.co/openai/whisper-large-v3}}, WavLM Large\footnote{\url{https://huggingface.co/microsoft/wavlm-large}}, HuBERT Large\footnote{\url{https://huggingface.co/facebook/hubert-large-ll60k}} and XEUS\footnote{\url{https://huggingface.co/espnet/xeus}}. All models except XEUS were initialized using Hugging Face checkpoints, while XEUS was initialized with the ESPNet Toolkit~\cite{watanabe18_interspeech}.

For prototyping efficiency, we pre-extracted features from the last hidden layer of each SSL model. This allowed us to rapidly iterate and compare their performances on the filtered validation set (with `X' and `O' labels removed). The model yielding the highest macro F1-score on this unimodal configuration was selected as the audio backbone for subsequent multimodal experiments.

\subsection{Bimodal Fusion with Text Encoder} \label{sec:text_encoder}

After selecting the best-performing audio encoder, we incorporated a text modality to provide content-based features. We used RoBERTa Large\footnote{\url{https://huggingface.co/FacebookAI/roberta-large}} as the text encoder, with transcriptions obtained with the state-of-the-art Canary ASR model~\cite{puvvada24_interspeech}. To ensure consistency, we relied solely on Canary-generated transcriptions, bypassing the dataset’s original annotations. RoBERTa embeddings were pre-extracted for efficiency.

Following, we explored multiple fusion strategies for integrating speech and text representations. A simple concatenation approach was first tested, where mean-pooled features from both modalities were combined and passed through an MLP classifier. Next, we evaluated early fusion via a single-layer Transformer encoder, refining each modality’s representation before applying mean pooling and classification~\cite{chen24_odyssey}. 

Additionally, we implemented Hierarchical Cross Attention (HCAM)~\cite{dutta2023hcam}, where bidirectional GRUs with self-attention refine SSL-based features before cross-attention layers enable inter-modal communication. Unlike~\cite{dutta2023hcam}, we applied an attentive pooling strategy before classification.

Finally, we examined the Multimodal Dual Attention Transformer (MDAT)~\cite{zaidi2024enhancing}, which integrates graph attention modules~\cite{velivckovic2018graph} and cross-attention mechanisms. MDAT utilizes two Transformer encoders for further refinement, differing from our early fusion setup by employing eight multihead attention heads, which yielded better performance in preliminary experiments.

\subsection{Prosodic and Spectral Feature Integration}

This work incorporates a prosodic modality derived from the F0, as previous work has shown that it is closely linked to speech emotion~\cite{Stasiak2012emotionf0}. We adopt a quantization-based approach, inspired by~\cite{wang2018autoregressive}, to model F0 contours. Raw F0, extracted using the RMVPE~\cite{wei23b_interspeech} model, is mel-scaled and quantized into 256 bins, with an additional padding index. These bins are mapped to learnable 256-dimensional embeddings, projected to 512 dimensions, and mean-pooled in the time dimension. This representation is concatenated with speech and text embeddings, improving the model context. To assess the efficacy of this quantized F0 embedding strategy, we conduct a comparative analysis against a baseline approach employing a 1D Convolutional Neural Network (CNN). This baseline CNN directly processes the raw F0 signal, utilizing a kernel size of 3, a stride of 1, and 256 channels. The CNN output is processed by a batch normalization layer followed by a ReLU activation function, and a linear projection layer that maps the output to a 512-dimensional space. Finally, mean-pooling is applied across the time dimension to generate a fixed-length representation. This comparison can be seen in Table~\ref{tab:combined}.

For spectral refinement, we extract Mel filterbanks via Kaldi-compatible torchaudio functions and process them with~\emph{Consistent Ensemble Distillation (CED)}~\cite{dinkel2024ced}. CED distills knowledge from large teacher models into lightweight yet high-performance student models. We employ the `CED-Small' variant (22M parameters), balancing computational efficiency and performance. The extracted spectral features are fine-tuned and concatenated with the fused embeddings before being passed to the MLP classifier, enriching the multimodal representation.

\subsection{SwiGLU Based MLP}

To further improve performance, we investigated the effect of replacing the standard ReLU activation with a SwiGLU in the MLP module. Inspired by~\cite{pacal2024enhancing}, we replaced the default MLP structure; consisting of a linear layer, ReLU activation, dropout, and a final classification layer with a SwiGLU-based module. The SwiGLU submodule comprises two linear layers and a Swish~\cite{ramachandran2018searching} activation, followed by a final linear layer. Our empirical results demonstrate that this modification yields a slight yet consistent performance improvement.


\section{Experimental Setup} \label{sec:exp}

In this section, we describe the experimental setup and the dataset utilized in our experiments, along with the data augmentation technique applied. 

All experiments were trained using a batch size of 8 and a Weighted Cross-Entropy Loss function, where class weights were determined by the inverse frequency of each class within the training set, for 20 epochs, using a cosine learning rate scheduler with warm-up of 500 steps. The learning rate was bounded between a minimum of 5e-5 and a maximum of 1e-5. Optimization was performed using the AdamW~\cite{loshchilov2018decoupled} optimizer, configured with the following parameters: $\epsilon$ = 1e-8, $\beta$ = {[}0.9, 0.98{]} and a weight decay of 1e-6. Additionally, norm-based gradient clipping, with a threshold of 10, was employed to stabilize training. 

\subsection{The MSP Dataset}  \label{sec:dataset}

The dataset used in this study is the MSP-Podcast corpus, a repository of spontaneous and emotionally diverse speech segments collected from various podcast recordings. Unlike traditional speech emotion recognition (SER) datasets, which often consist of acted emotional expressions, this corpus captures natural interactions, providing greater variability and authenticity.

\subsection{Data Augmentation}

We enhance generalization and mitigate overfitting using SeqAug~\cite{georgiou2023seqaug}, a modality-agnostic sequential data augmentation technique applicable to our pre-extracted speech and text features. SeqAug permutes randomly selected feature dimensions temporally, resampling from a distribution while preserving sequence integrity. During training, we apply SeqAug independently to each modality with 50\% probability, consistently using a beta distribution ($\alpha=0.5$) for permutation. Unlike the original, our approach employs independent permutations per feature dimension, rather than a shared permutation to force the model to learn representations robust to temporal misalignments between features, simulating asynchrony between modalities and improving feature importance extraction.




\section{Results and Discussion} \label{sec:res}

In this section, we analyze the performance of our systems, evaluating different aspects of feature extraction and fusion strategies. We report results for unimodal speech models, bimodal fusion with text, prosodic and spectral feature integration. Finally, we present the performance of the ensemble system, which represents our best configuration.

Performance was measured using the following metrics: Macro and Micro F1-score, Recall (R), and Precision (P). For brevity, we did not report accuracy since Micro F1 and accuracy yielded identical values.

\begin{table}[ht] \footnotesize \centering
\caption{Performance metrics for unimodal speech models. Best results are in bold, and the second-best results are underlined.}
\label{tab:ssl} 
\begin{tabular}{l|c|c|c|c}
\hline
\textbf{SSL Model} & \multicolumn{1}{l|}{\textbf{Macro F1}} & \multicolumn{1}{l|}{\textbf{Micro F1}} & \multicolumn{1}{l|}{\textbf{R}} & \multicolumn{1}{l}{\textbf{P}} \\ \hline
W2V2 Large     & 0.178                                  & 0.394                             &   0.198       &     0.172       \\ \hline
HuBERT Large   & 0.274                                  & 0.478                             & 0.279                           & 0.276                           \\ \hline
WavLM Large    & 0.313                                  & 0.482                             & 0.333                           & 0.316                           \\ \hline
Whisper Large V3     & \textbf{0.366}                                  & \textbf{0.524}                             & \textbf{0.391}                           & \textbf{0.366}                           \\ \hline
XEUS           & \underline{0.323}                                 & \underline{0.512}                             & \underline{0.334}                           & \underline{0.323}                           \\ \hline
\end{tabular}
\end{table}

\begin{table*}[!ht] \centering  \footnotesize
\caption{Comparison of the performance of the individual models selected to build the ensemble and the overall performance of the ensemble system. Best results are in bold, and the second-best results are underlined. "Batch Bal" refers to the usage of balaned sampling as in~\cite{kong2020panns}, "Focal" refers to the usage of Focal Loss~\cite{lin2017focal}.}
\label{tab:ensemble}
\begin{tabular}{l|c|c|c|c}
\hline
\textbf{Model}      & \textbf{Macro F1} & \textbf{Micro F1} & \textbf{R} & \textbf{P} \\ \hline
Whisper + RoBERTa + MDAT + F0 Quant + SwiGLU + Data Aug & \underline{0.411} & \underline{0.567} & 0.420 & \textbf{0.446} \\
Whisper + RoBERTa + HCAM + F0 Quant + CED Spectral & 0.367 & 0.496 & \underline{0.431} & 0.385 \\
Whisper + RoBERTa + HCAM + F0 Quant + CED Spectral + Data Aug + Focal & 0.384 & 0.547 & 0.397 & 0.386 \\
XEUS + E5~\cite{wang2024multilingual} + Transformers + Batch Bal + Data Aug + Focal & 0.335 & 0.505 & 0.353 & 0.337 \\
\hline
Ensemble (Majority Voting) & \textbf{0.422} & \textbf{0.581} & \textbf{0.437} & \underline{0.414} \\
\hline
\end{tabular}
\end{table*}

\begin{table}[ht]
\footnotesize
\caption{Performance of various fusion strategies, prosodic and spectral feature integration. Best results within each section are in bold, and the second-best results are underlined.}
\label{tab:combined}
\centering
\begin{tabular}{l|r|r|r|r}
\hline
\textbf{Strategy/Feature}      & \textbf{Macro F1} & \textbf{Micro F1} & \textbf{R} & \textbf{P} \\ \hline
\multicolumn{5}{l}{\textbf{Fusion Strategies}} \\ \hline
Simple                         & \underline{0.388}          & \underline{0.557}        & \underline{0.404}      & 0.383      \\
Transformers                   & 0.364          & 0.448        & 0.362      & 0.377      \\
HCAM                           & 0.383          & 0.556        & 0.391      & \underline{0.396}      \\
MDAT                           & \textbf{0.401}          & \textbf{0.582}        & \textbf{0.393}      & \textbf{0.455}      \\ 
\hline
\multicolumn{5}{l}{\textbf{Prosody Integration}} \\ \hline
+ F0                          & 0.397          &  \textbf{0.587}    &    0.398   &   \underline{0.435}    \\
+ Quant. F0                                    & \underline{0.407}          & 0.572        &  \underline{0.410}      & 0.425      \\
\hspace{0.5em} + Data Aug.                    & 0.400          & \underline{0.586}        & 0.400      & 0.436      \\
\hspace{1.0em} + SwiGLU                       & \textbf{0.411}          & 0.567        & \textbf{0.420}      & \textbf{0.446}      \\ 
\hline
\multicolumn{5}{l}{\textbf{Spectral Integration (CED)}} \\ \hline
\hspace{0.5em} + Random                              & 0.342         & \underline{0.583}        & 0.332      & \textbf{0.464}     \\
\hspace{0.5em} + Pretrained                    &0.376           &0.578        &0.366      & 0.403       \\ 
\hspace{1.0em} + Data Aug.                    &\underline{0.393}           &0.568         &\underline{0.395}       &0.409      \\
\hspace{1.5em} + SwiGLU & \textbf{0.405}          & \textbf{0.586}        & \textbf{0.396}      & \underline{0.435}      \\  \hline
\end{tabular}
\end{table}

Table~\ref{tab:ssl} shows unimodal speech model performance.  Wav2Vec2 and HuBERT underperformed, likely due to pretraining on read speech.  WavLM, XEUS, and Whisper performed better, with Whisper achieving the highest Macro F1 (0.366) and Micro F1 (0.524), likely benefiting from pretraining on diverse, spontaneous, and noisy speech.


Table~\ref{tab:combined} presents results for bimodal fusion (incorporating RoBERTa Large text features), comparing multiple fusion strategies.  Bimodal fusion significantly outperformed unimodal models.  MDAT, integrating GATs, achieved the highest Macro F1 (0.401), highlighting the effectiveness of graph-based multimodal fusion.  Surprisingly, the simple fusion strategy outperformed more complex techniques like Transformer-based fusion and HCAM.  This may be due to increased susceptibility to overfitting with the more complex models, given the limited dataset size.



Integrating pretrained CED spectral features with prosodic features, data augmentation, and SwiGLU yielded a Macro F1 of 0.405, demonstrating the value of incorporating learned spectral representations in addition to the prosodic information. The best-performing CED and F0 configurations performed similarly (0.405 vs. 0.411), suggesting both feature types contribute valuable, complementary information. Notably, a model employing randomly initialized CED features performed significantly worse (Macro F1: 0.342), highlighting the importance of pre-training, which performed better (Macro F1: 0.376). A SwiGLU-based MLP consistently improved performance, particularly when used alongside SeqAug, indicating its effectiveness in capturing temporal relationships.

Our speech emotion recognition ensemble was built using an exhaustive search algorithm, exploring various model combinations. This exploration encompassed 13 variations of the top-performing configuration. These variations experimented with different fusion methods, the inclusion of data augmentation, the addition of prosodic and spectral features, and the integration of the second-best performing unimodal model (XEUS) and an alternative text encoder (E5 Large V2). We also tested using focal loss and balanced sampling. While E5 demonstrated strong performance on standard text benchmarks, initial testing revealed its lower effectiveness compared to RoBERTa Large in this specific emotion recognition task. Ensemble candidates, always including our best individual model, using at least three models at a time.

Predictions were combined via majority voting, with our top model breaking ties. To prevent overfitting, ensembles were evaluated on 100 randomly sampled, class-balanced subsets of the validation data. The size of each subset equalled the smallest class sample size. The final ensemble was selected based on the highest mean macro-F1 score across these subsets.

As presented in Table~\ref{tab:ensemble} our final ensemble achieved a Macro F1-score of 0.422, representing an improvement over individual configurations, including the best-performing single model (Whisper with a Macro F1 of 0.366) and the most effective fusion strategy (MDAT, with a Macro F1 of 0.401 and 0.411 with the usage of data augmentation and SwiGLU). This significant increase indicates that the combination of diverse model variations, along with complementary prosodic and spectral features, effectively boosted the system’s robustness and its ability to generalize across class imbalances and subtle emotional expressions in spontaneous speech.

\section{Conclusion} \label{sec:conc}

This work presented a multimodal system for speech emotion recognition in naturalistic conditions, leveraging SSL-based speech models, a text encoder, and prosodic and spectral features. Our evaluation of unimodal models demonstrated the strong performance of Whisper and XEUS, highlighting their robustness for SER in spontaneous speech.

Among fusion strategies, the Multimodal Dual Attention Transformer (MDAT) achieved the best results, confirming the effectiveness of graph-based attention. Additionally, the integration of quantized F0 embeddings significantly improved Macro F1, reinforcing the importance of prosodic cues. Our final ensemble model reached a Macro F1-score of 0.422, demonstrating the advantages of combining multiple modalities and advanced fusion techniques.

In future work, we plan to explore additional fusion strategies to further enhance cross-modal interactions and capture nuanced emotional expressions. We also aim to refine our data augmentation and ensemble selection methods, investigating end-to-end training frameworks that may reduce computational overhead while improving model interpretability and generalization. Furthermore, validating our approach on larger, more diverse datasets and incorporating domain adaptation techniques will be critical to ensuring robustness across different speech contexts and application scenarios.

\section{Acknowledgements}

This work has been fully/partially funded by the project Research and Development of Algorithms for Construction of Digital Human Technological Components supported by Advanced Knowledge Center in Immersive Technologies (AKCIT), with financial resources from the PPI IoT/Manufatura 4.0 / PPI HardwareBR of the MCTI grant number 057/2023, signed with EMBRAPII/. We also thank the Artificial Intelligence Lab at Recod.ai, the Institute of Computing, University of Campinas.

\bibliographystyle{IEEEtran}
\bibliography{mybib}

\end{document}